\documentclass[twocolumn,showpacs, numbers,amsmath,amssymb,prl]{revtex4}
\usepackage{graphicx}
\usepackage{amsmath, amsthm, amssymb,mathtools}
\usepackage{latexsym}
\usepackage{amssymb}
\usepackage{bm}
\usepackage{dcolumn}
\usepackage{epstopdf}

\usepackage{color}

\newcommand{\Ox}{$^{17}$O}
\newcommand{\Cu}{$^{63}$Cu}
\newcommand{\Hg}{$^{199}$Hg}

\newcommand{\YBCO}{YBa$_2$Cu$_3$O$_{7-\delta}$}
\newcommand{\BSCCO}{Bi$_2$Sr$_2$CaCu$_2$O$_{8+\delta}$}

\newcommand{\HBCO}{HgBa$_2$CuO$_{4+\delta}$}
\newcommand{\LSCO}{La$_{2-x}$Sr$_x$CuO$_4$}
\newcommand{\Hparac}{$H_0\,||\,c$}

\newcommand{\Hparaa}{$H_0\,||\, a $}
\newcommand{\Hperpc}{$H_0\perp c$}
\newcommand{\Tc}{$T_c$}
\newcommand{\degrees}{$^{\circ}$}
 
\begin{document}

\title{Absence of static orbital current magnetism at the apical oxygen site in \HBCO\ from NMR}
\author{A.M. Mounce$^1$, Sangwon Oh$^1$, Jeongseop A. Lee$^1$, W.P. Halperin$^1$, A.P. Reyes$^2$, P.L. Kuhns$^2$, M.K. Chan$^3$, C. Dorow$^3$, L. Ji$^3$, D. Xia$^{3,4}$, X. Zhao$^{3,4}$, and M. Greven$^3$ }
\affiliation{$^1$Department of Physics and Astronomy, Northwestern  University, Evanston, IL 60208, USA\\
$^2$National High Magnetic Field Laboratory, Tallahassee, FL 32310, USA\\
$^3$School of Physics and Astronomy, University of Minnesota, Minneapolis, Minnesota 55455, USA\\
$^4$College of Chemistry, Jilin University, Changchun 130012, China}

\date{Version \today}

\pacs{74.25.nj, 74.72.Kf, 76.60.Jx}

\begin{abstract}
The simple structure of \HBCO\ (Hg1201) is ideal among cuprates for study of the pseudogap phase as a broken symmetry state. We have performed $^{17}$O nuclear magnetic resonance (NMR) on an underdoped Hg1201 crystal with transition temperature of 74 K to look for circulating orbital currents proposed theoretically and inferred from neutron scattering. The narrow spectra preclude static local fields in the pseudogap phase at the apical site, suggesting that the moments observed with neutrons are fluctuating. The NMR frequency shifts are consistent with a dipolar field from the Cu$^{+2}$ site.

\end{abstract}

\maketitle

A distinguishing feature of cuprate superconductors, other than their high transition temperature, is the existence of a normal state pseudogap in underdoped materials. However, the origin of the pseudogap is not yet understood.  There are many suggestions that it is consistent with a broken symmetry state such as staggered fluctuating currents\cite{wen96}, orbital currents which conserve translational symmetry\cite{var99, sim02, var06}, $d$-density wave\cite{cha01}, and various other broken symmetries\cite{eme97, kiv98}. 

Polarized neutron scattering in \YBCO\cite{fau06,sid07,moo08}, \HBCO\ (Hg1201)\cite{li08, li11}, \BSCCO \cite{dea12}, and \LSCO\ (short ranged)\cite{bou11} show the appearance of broken time reversal symmetry that correlates with the onset of the pseudogap phase with a magnetic moment tilted away from the crystalline $c$-axis. A muon spin relaxation ($\mu$SR) experiment in zero field was unable to detect this magnetism\cite{mac08} although screening effects may account for muon insensitivity.\cite{she08} To account for neutron data, extensions of 2D orbital currents to  3D models  have been proposed which include the apical oxygen\cite{web08} as well as a quantum mechanical calculation \cite{he12} which has an out-of-plane component with a tilted local field.  In this context, magnetic fields at the apical oxygen site, Fig.1(a), up to 200 G have been calculated which should be measurable by NMR.\cite{led12}

Both NMR and nuclear quadrupolar resonance (NQR) experiments have previously attempted measurement of orbital currents: $^{89}$Y NMR in Y$_2$Ba$_4$Cu$_7$O$_{15+\delta}$\cite{str08} and $^{135}$Ba NQR in YBa$_2$Cu$_4$O$_8$.\cite{str11}  However, $^{89}$Y is in a symmetry position which is insensitive to orbital current ordering and neutron data is not available in these materials for direct comparison. We report here our investigation of local fields in Hg1201 with \Ox\ NMR for which the apical oxygen is sensitive to orbital currents and polarized neutron scattering show evidence for magnetic ordering.  

\begin{figure}[b!]
\begin{center}
	\includegraphics[width=.45\textwidth]{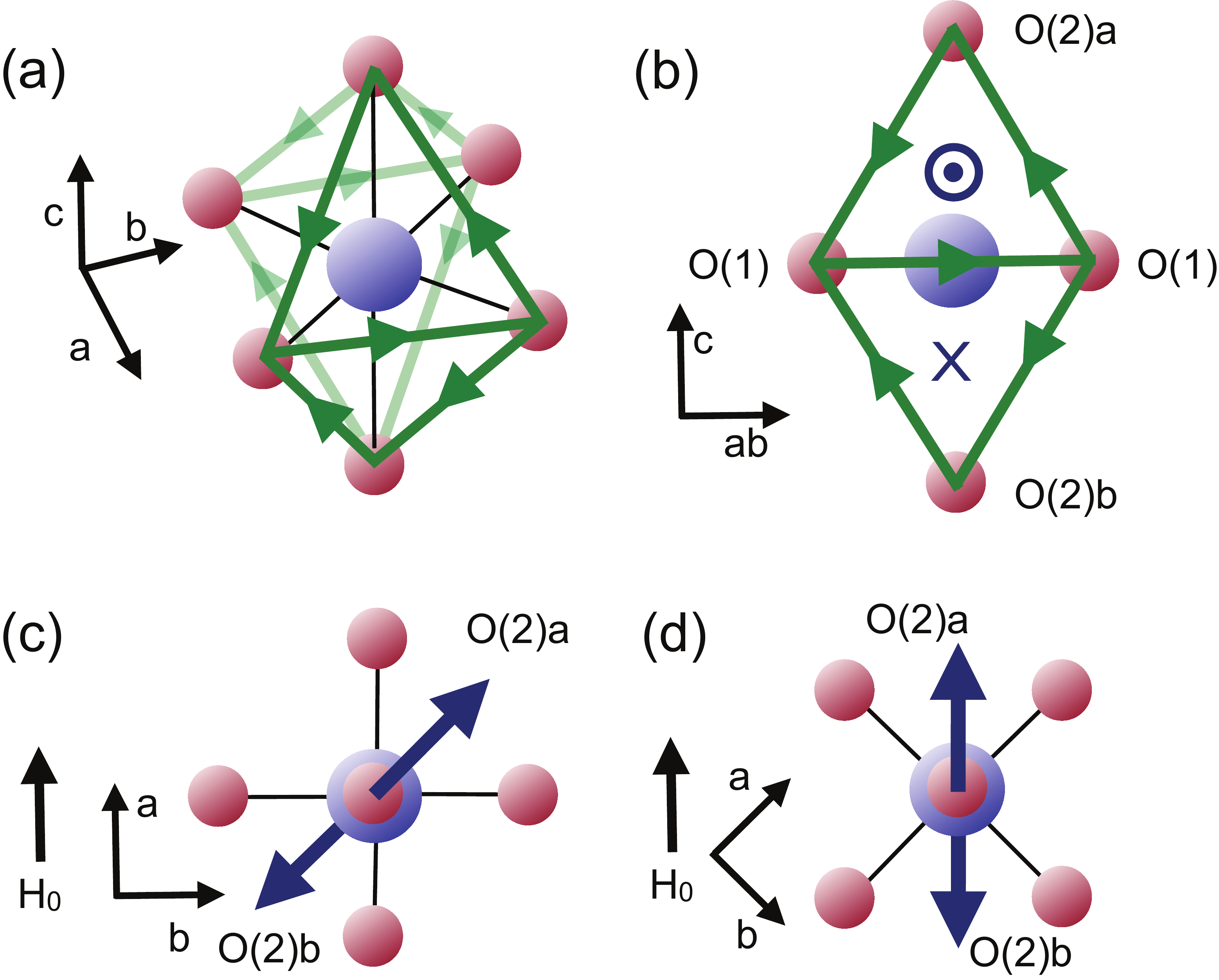}
	\caption{(a) Orbital current ordering (green arrows) in the CuO$_6$ octahedron in Hg1201 proposed~\cite{web08,led12} as explanation for neutron scattering results. (b) The profile of the currents and the effective magnetic fields (blue symbols) produced. The apical oxygens will experience different field depending on lattice position with O(2)a above and O(2)b below the CuO$_2$ plane while the local fields will cancel at the O(1) site by symmetry. (c) \& (d) A view of the CuO-plane along the $c$-axis, showing the local fields (blue arrows) at the apical oxygen (above and below the Cu nucleus) at two different orientations: \Hparaa\ and $H_0= 45$\degrees\ between the $a$ and $b$-axes.}
\label{orbitalcurrents}
\end{center}
\end{figure}

Until recently, NMR results for Hg1201  have been constrained to $c$-axis magnetically aligned powdered samples,\cite{suh96a, bob97,gip97,gip99}  which are more easily annealed and have a large NMR signal. However, these powder samples are disordered in the $ab$-plane and they can have a larger spectral linewidth due to imperfect alignment. In order to investigate possible effects of orbital currents, high quality single crystals are necessary.  These samples have become available\cite{zha06, bar08} and the results of \Hg\ and \Cu\ NMR\cite{ryb09, haa12} have  been reported. We isotope exchanged \Ox\ and performed high resolution NMR frequency shift measurements for both the planar O(1) and apical O(2) oxygen, with narrow linewidths of approximately 12 kHz and 5kHz respectively at $H_0$ = 6.4 T, for a moderately underdoped sample with superconducting transition temperature of 74 K, as compared to 95 K at optimal doping.

If the pseudogap state breaks time reversal symmetry from orbital currents, as shown in Fig.1(a), the two apical oxygens in each unit cell will experience oppositely directed  field components, Fig.1(b).  Furthermore, the orientation of these magnetic fields are locked at 45\degrees\,  from the $a$ or $b$-axes.  These field components have an angular dependence with a maximum value, $H_{max}$, between the $a$ and {$b$-axes}, $H_{max}/\sqrt{2}$ along each axis, and zero parallel to the $c$-axis.  With applied magnetic field in the $ab$-plane the orbital currents will produce a splitting of the central transition with a period of $\pi$ for rotation about the $c$-axis.

\begin{figure}[t!]
\begin{center}
		\includegraphics[width=.45\textwidth]{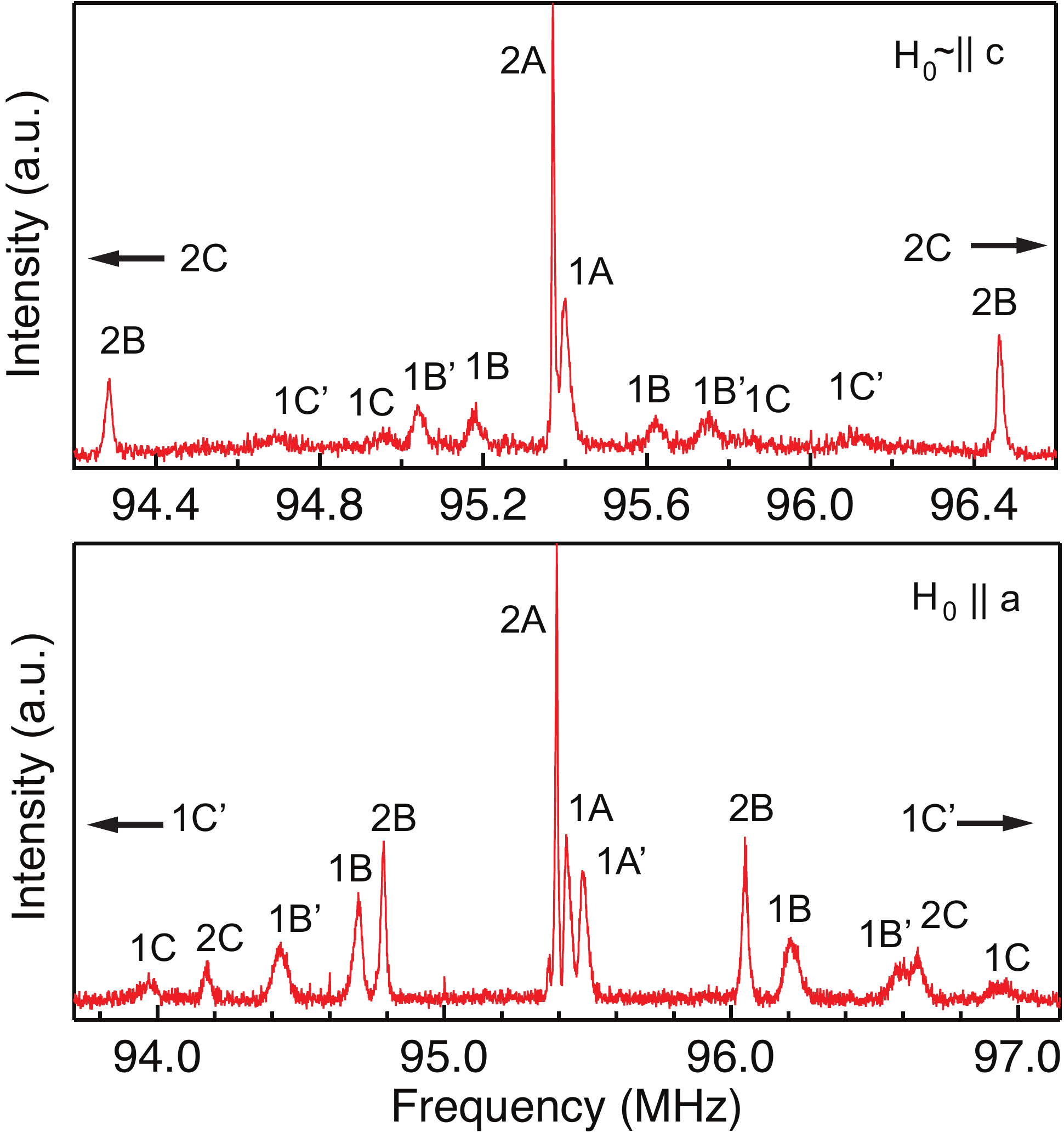}

		\caption{NMR spectra at $H_0$ = 16.5 T and $T = $ 100 K, including satellite transitions, for \Hparac\ and \Hparaa. The transitions are labeled $n\alpha$ where $n = 1, 2$ for O(1), O(2) and $\alpha$ = A, B, C
 for the central, first quadrupolar, and second quadrupolar transitions. The central transition and four satellites can be seen for three distinct oxygen sites, apical,  planar with Cu-O bond along the field, and  planar with the Cu-O bond perpendicular to the field (primed).  The planar sites should be degenerate with \Hparac\ however the crystal is misaligned by $\approx$10 degrees, evident by two sets of satellites in this orientation. The higher order satellites 2C and 1C' are omitted for clarity.}
\label{spectra}
\end{center}
\end{figure}

\begin{figure}[t!]
\begin{center}
		\includegraphics[width=.45\textwidth]{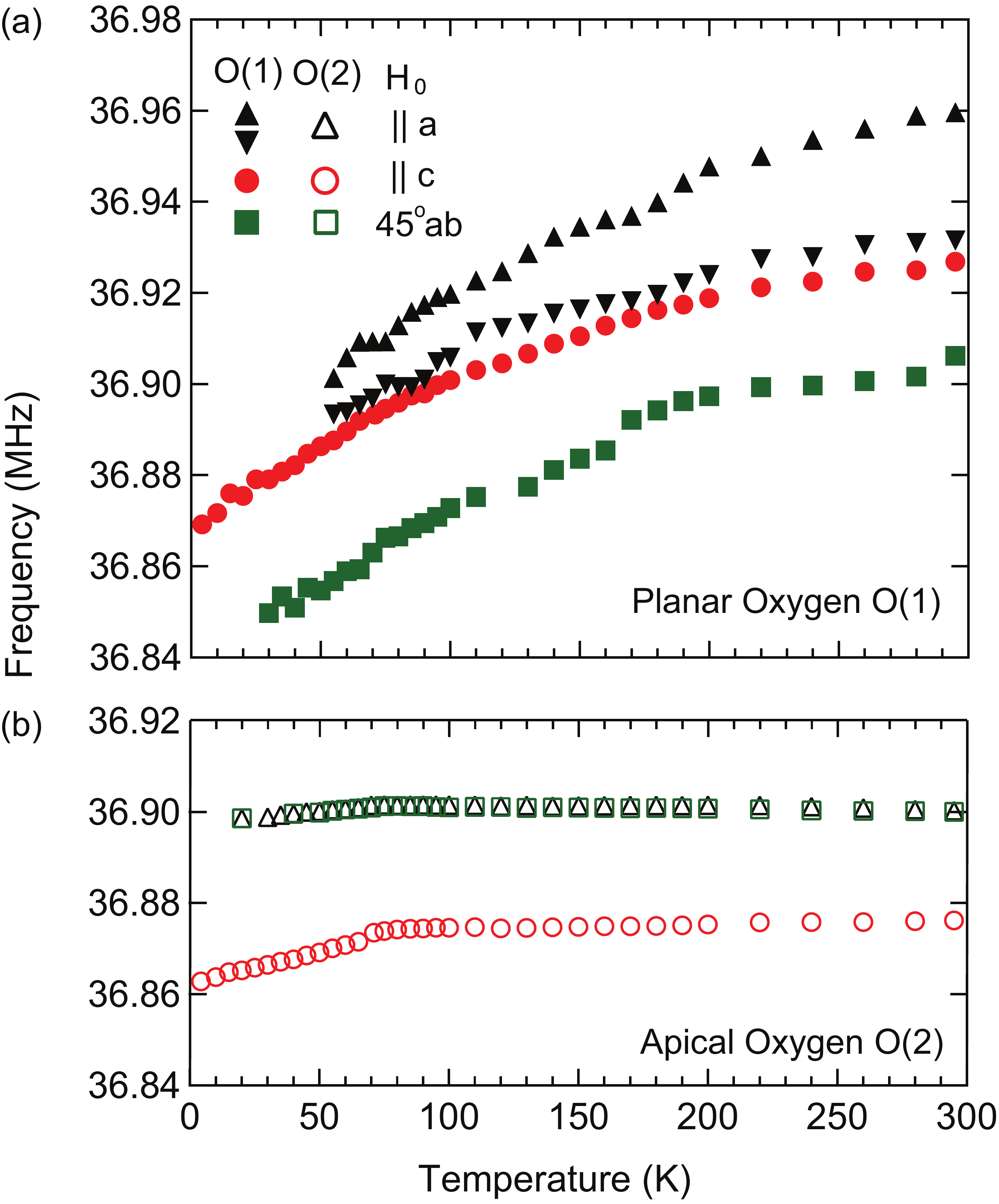}

	\caption{Temperature dependence of the planar O(1) and apical O(2) oxygen shifts with \Hparac, \Hperpc\  with $H_0\,||\,a$ and with $H_0$ at 45\degrees\  to the $a$ and $b$-axes.  For \Hparaa, there are two inequivalent O(1) sites.\cite{cro11} The O(1) shift shows the classic decrease of $K_s$ characteristic of the pseudogap.  The O(2) shift has a minimal $K_s$ component which allows for clear observation of the reversible diamagnetism in the superconducting state.}
\label{tempshifts}
\end{center}
\end{figure}

Single crystals were grown by self flux method\cite{zha06,bar08} then heat treated for \Ox\ exchange  and  annealing.  A 20 mg sample was held  at 600\degrees C for 12 hours in 1 atm of 90\%  \Ox\  and then, in a continuous step, lowered to 300\degrees C for 10 days.  The annealing resulted in an underdoped superconducting transition temperature,  \Tc , of 74.0 K with a transition width, $\Delta T_c$, of 3 K, compared to 74.5 K and 3 K, respectively, before exchange.

Nuclear magnetic resonance was performed in fields of $H_0$ = 6.4 and 16.5 T and temperatures from 4.2 to 295 K. Both stoichiometric sites were readily observed and the resonant frequency was taken from the peak of the spectra. Spectra were measured by a standard Hahn echo sequence, $\pi/2 - \tau - \pi$, with a typical $\pi /2$ pulse length of 2 $\mu$s, sufficient to excite both sites. For full spectra, including satellite transitions, Fig.2, a frequency sweep method was used. From the full spectra the quadrupolar splitting was found to be $\nu_Q$ = 1.08 MHz for both sites with the axis of symmetry along the $a$ and $c$-axes for  O(1) and O(2) respectively.  The O(2) site exhibits axial symmetry while the O(1) site has less than axial symmetry with an asymmetry parameter $\eta = 0.40$.  With \Hperpc, we were able to separate the O(1) and O(2) resonances taking advantage of their rather different spin-spin relaxation times, i.e. O(2) has a much longer relaxation time and consequently, for sufficiently long $\tau$, O(1) loses phase coherence, leaving only an O(2) contribution to the signal amplitude.

The NMR spectrum frequency shift $\nu-\gamma H_0 = K(T,H_0)\gamma H_0$, where $\nu$ is the resonant frequency, $\gamma$ is the gyromagnetic ratio, and $K(T,H_0)$ can be written as a sum of components such that,
\begin{equation}
 K(T, H_0)= K_{s}(T) +K_{o}+K_{Q}+4\pi(1-D) M(T)/H_0,
\end{equation}
\noindent
and $K_s(T)$ is the spin shift, called the Knight shift; $K_o$  is the orbital shift; $K_{Q}$ is the quadrupolar shift; $D$ is the demagnetization factor (dependent on sample geometry), and $M(T)$ is the magnetization due to superconducting screening currents,  all of which can be anisotropic, although only $K_s(T)$ and $M(T)$ have temperature dependence. As is usually considered for cuprates,\cite{mil89b, sha89} we adopt a single-component spin susceptibility model, $K_s = A \chi_s (T)$, where $A$ is the transferred hyperfine field for a particular nucleus and $\chi_s(T)$ is the electronic spin susceptibility.

Our main results are the orientation and temperature dependence of the O(1) and O(2) frequency shifts of the central transitions, Fig.3. For all orientations, the O(1)  shift exhibits the classic manifestation of the pseudogap\cite{all89, tim99}, decreasing from above room temperature down to our minimum temperature $T = 4$ K with no clear indication of the transition to the superconducting state. In our sample, onset of the pseudogap phase can be estimated from polarized neutron scattering\cite{li08} to be at $T\approx 270$ K. The shift of O(2) behaves significantly differently.  Above \Tc\ there is little temperature dependence, while below \Tc\ there is a sharp onset for a decreasing temperature dependent shift which we attribute to superconducting diamagnetism.

\begin{figure}[t!]
\begin{center}
		\includegraphics[width=.45\textwidth]{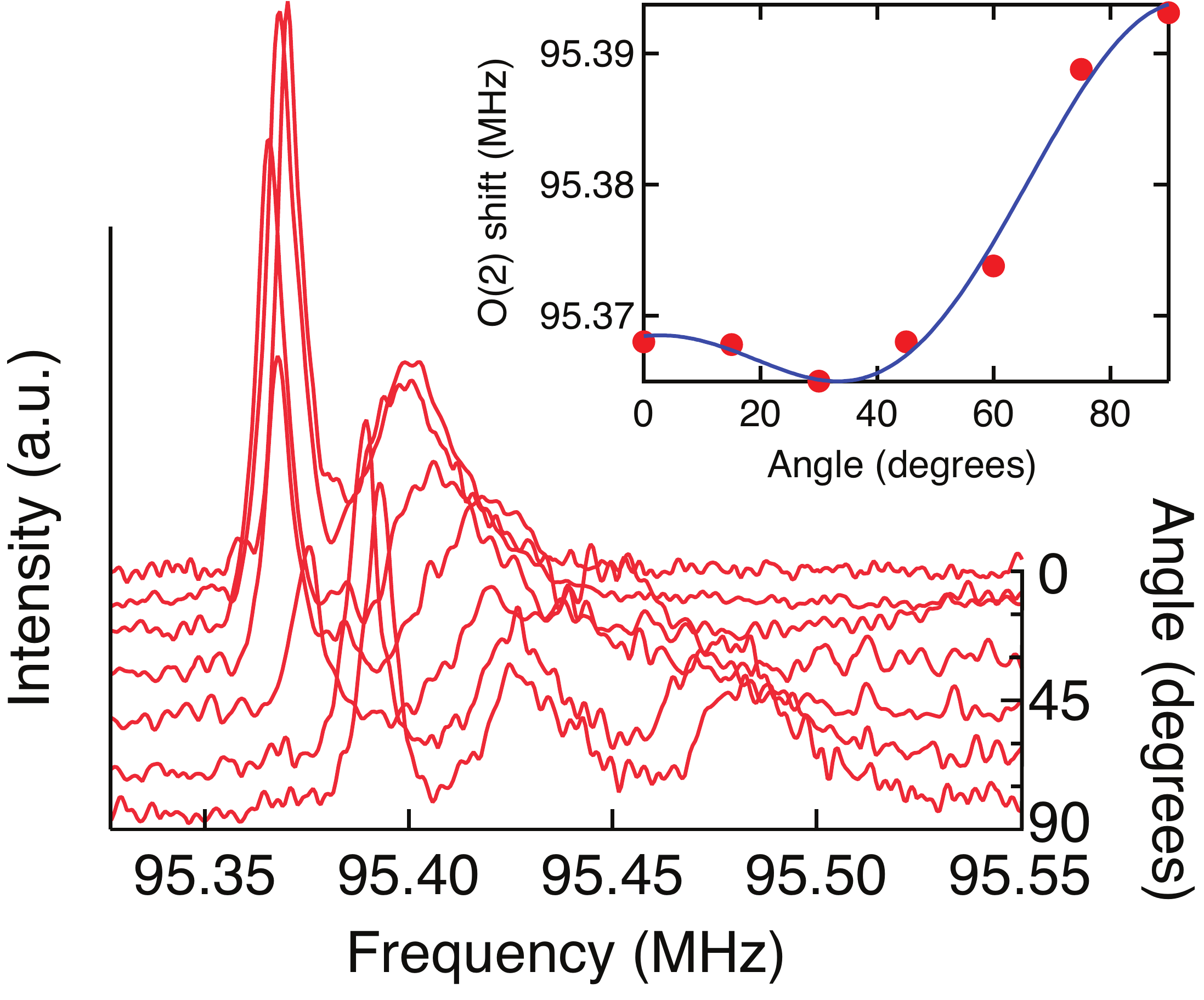}

	\caption{Central transitions at T = 100 K at 15\degrees\ intervals from $\theta$ = 0 to 90\degrees, or \Hparac\ to \Hparaa\ respectively at $H_0=16.5$ T.  For the planar oxygen, the angular dependence shows the rhombic symmetry of the orbital, spin, and second order quadrupolar shift. The apical oxygen has axial symmetry along the $c$-axis.  Inset: Angular dependence of the O(2) second order quadrupolar and orbital shifts, Eq.2, given by the blue curve. }
\label{angularspectra}
\end{center}
\end{figure}

\begin{figure}[blp!]
\begin{center}
		\includegraphics[width=.45\textwidth]{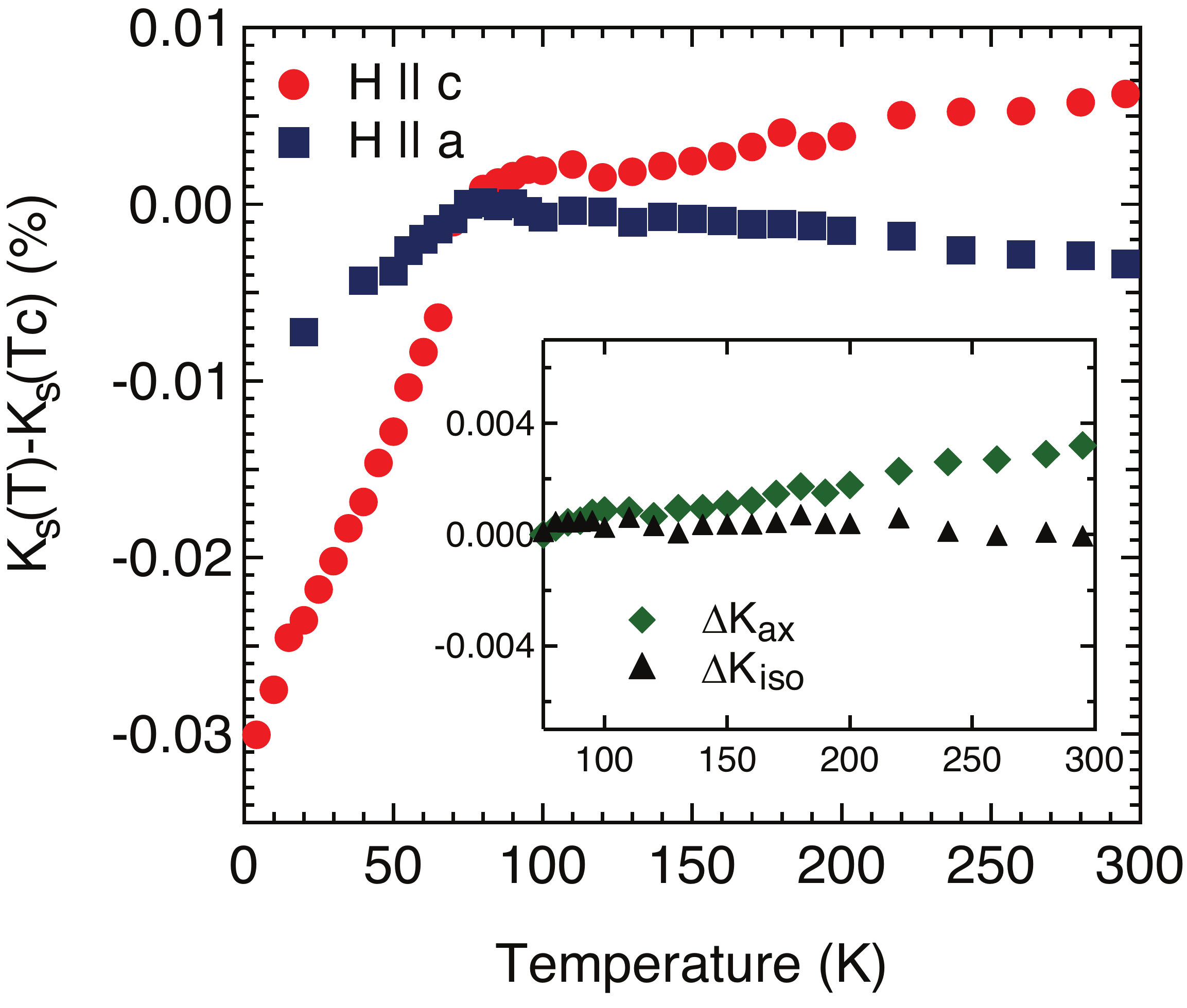}
\caption{Frequency shifts of the apical oxygen in the underdoped sample with orientations \Hparac\ and \Hparaa\ as a function of temperature relative to the value at \Tc . The shift below \Tc\ can be attributed to  superconducting diamagnetism.  Inset: Above \Tc\ the shifts are deconvolved into isotropic and axial components showing that only $\Delta K_{ax}$ is temperature dependent. }
\label{shiftsc}
\end{center}
\end{figure}

We look for the existence of orbital current order, contributing to $K_{o}$, in two different ways.  First, we compare the field dependence of the shift along the $c$-axis with that in  the $ab$-plane along the $a$-axis.  Secondly, we measure the angular dependence of the shift and the linewidth for a magnetic field within the $ab$-plane. 

The spin susceptibility is a strong function of temperature, evident from the O(1) shift, decreasing by over half its value from room temperature down to  \Tc, Fig.\ref{tempshifts}.  Thus, the small temperature dependence of the O(2) shift in the normal state indicates that the transferred hyperfine field from the CuO-plane is correspondingly small. Therefore,  the major contributions to the shift of the apical oxygen can be taken to be,
\begin{equation}
 K_{O(2)}(\theta, H_0) = K_o(\theta) + K_Q^{(2)}(\theta, H_0),
\end{equation}
\noindent
where $\theta$ is the angle between the magnetic field and the $c$-axis.  The central transition is unaffected by the quadrupolar interaction to first order, so $K_Q^{(2)}$ is the second order shift given by the blue curve in the inset to Fig.\ref{angularspectra}: $\nu = \nu_Q^2/2\nu_0(1-9cos^2\theta)(3cos^2\theta-1)$, where $\nu_0$ is the Lamor frequency and  $\nu_Q$ is the quadrupolar splitting from the satellite spectra, Fig.\ref{spectra}. Our data for O(2) closely follow this behavior, which after subtraction leaves the orbital components.  Circulating orbital currents would shift the resonance frequency an amount independent of applied field.  At $T =$ 100 K, and $H_0=6.4  \,(16.5)$ T,  we find the difference $\Delta K_o=K_o(\theta=90^{\circ})-K_o(\theta=0^{\circ}) =  0.025$\% \,(0.027)\%, to be almost field independent placing a limit on a possible excess field from orbital currents of 2.1 G in the $ab$-plane.

For an independent estimate of the effect of orbital currents in the $ab$-plane, we have measured the angular dependence of the shifts at a magnetic field of $H_0 = 6.4$ T rotated about the $c$-axis. As can be seen from Fig.3(b), the apical oxygen NMR frequency has no in-plane angular dependence  within our resolution of 2 G, indicating that there is little, or no, additional contribution from orbital currents.  A more stringent test, if the orbital currents have the symmetry shown in Fig.\ref{orbitalcurrents}, is to examine the NMR linewidth for broadening that might reflect a small frequency splitting.  Our measurements at all temperatures of the angular variation of the linewidth is less than 0.5 kHz which places a bound on  orbital current fields of 0.4 G. According to Lederer and Kivelson,\cite{led12} orbital currents inferred from neutron scattering results\cite{fau06, moo08, li08} should lead to local fields of order $H_{max}\sim 200$ G oriented as indicated in Fig.\ref{orbitalcurrents}.  Taken together, our measurements of the angular variations of the apical orbital shift and NMR linewidth, in-plane and out of-plane, place substantial constraints on predictions from the theory for a static orbital field to be less than 0.4 G.

Examining the O(2) shift in greater detail reveals that the O(2) nucleus has a small temperature dependent spin shift in the normal state such that  from room temperature to \Tc , $\Delta K_s$ equals $(-0.0035, -0.0035, 0.0065)$\% in the crystal basis $(a,b,c)$, Fig.\ref{shiftsc}. The temperature dependence of the shifts can be deconvolved into isotropic, $K_{iso}$, and axial, $K_{ax}$, components where the latter has the angular dependence of the second order Legendre polynomial.  For O(2), the shift is axially symmetric around the $c$-axis and the shifts are given by,

\begin{equation}
  \begin{array}{l l}
   	$$K_{iso} = (K_c + 2K_{ab})/3$$ \\
	$$K_{ax} = (K_c - K_{ab})/3$$
 \end{array} 
  \end{equation}

This decomposition shows that all of the temperature dependence resides in the axial component of the spin shift, Fig.\ref{shiftsc}.  The transferred hyperfine field from the Cu$^{+2}$ ion and contributions to the ligand O$^{-}$ hyperfine field, can come from several sources\cite{owe66}:

\begin{equation}
  \begin{array}{l l}
   	$$A_{ iso} = A_{c}+A_{cp}$$ \\
	$$A_{ax} =A_{p}+A_{dip}$$
 \end{array} 
  \end{equation}
  
\noindent
where $A_{iso}$ ($A_{ax}$) are the isotropic (axial) hyperfine fields, $A_{c}$ is the Fermi contact term, $A_{cp}$ is from core polarization, $A_{p}$ is from unpaired $p$ electrons (only $p$ is relevant for the O(2) nucleus, although $d$ and $f$ would also be axial), and  $A_{dip}$ is the direct dipolar coupling from spin on the Cu$^{+2}$ ion. From the temperature independence of $K_{iso}$ we infer that the hyperfine fields $A_{c}$ and $A_{cp}$ are negligibly small, corresponding to temperature dependent shifts less than 3\,x\,10$^{-4}\%$.  Core polarization comes from the unpaired $p$ electron and $A_{p} \propto A_{cp}$.  The ratio can be estimated as $A_{cp}/A_{p}= (10/4) \kappa$ where $\kappa$ is an empirical constant.  For atomic fluorine (the same electronic configuration as O$^-$), $\kappa = -0.1$\cite{har65}, giving a ratio of $A_{cp}/A_{p}=0.25$.  This would place an upper bound on $A_{p}$ corresponding to at most a 40\% contribution to $\Delta K_{ax}$, suggesting that the majority of its contribution comes from a dipolar field  such as a magnetic moment on the Cu site. \cite{tak93, suh96a} 

Presuming the dipolar effect at the O(2) site  also exists at the planar O(1) site, and taking the O(2) shift from 300 K to $T_c$, $\Delta K_{ax,O(2)} = \Delta K_{dip,O(2)} = 0.0035$\%, we find for the O(1) site $\Delta K_{dip, O(1)} = 0.021 \%$, noting that it has two Cu nearest neighbors and using known atom distances.\cite{put93}  This shift agrees quite well with our measured anisotropic shift of O(1), $\Delta K_{ax, O(1)}= 0.018\%$, indicating that $K_{ax,O(1)}$ has a significant dipolar contribution from the Cu$^{+2}$ site as well.

The hyperfine fields at the apical oxygen site are sufficiently small that we can readily observe the shift in the superconducting state from superconducting diamagnetism, Fig.\ref{shiftsc}.  Taking into consideration the demagnetization factor that is dependent on the overall crystal dimensions, we determine the superconducting penetration depths, $\lambda_\alpha$, where $\alpha = ||\ \mathrm{or} \perp$ denotes the orientation of the field relative to the $c$-axis, $\Delta K = K(T_c ) - K(T = 0) = 4 \pi (1-D_\alpha)M_\alpha/H_0 $.  The crystal dimensions give demagnetization factors $D_{\perp}$ = 0.67 and $D_{||}$ = 0.165 \cite{par04}. The magnetization, $M_\alpha (H)$, can be expressed by:
\begin{equation}
M_{\alpha}(T)= \left\{
  \begin{array}{l l}
    \frac{\Phi_0}{8\pi\lambda_{||}^2(T)}\ln\frac{H_{c2}}{H}, & \quad \text{for } \alpha =  \parallel \\
    \frac{\Phi_0}{8\pi\lambda_{||}(T)\lambda_{\perp}(T)}\ln\frac{H_{c2}}{H}, &  \quad \text{for } \alpha = \perp \\
 \end{array} \right.
 \end{equation}
\noindent
where $\Phi_0$ is the magnetic flux quantum and $H_{c2}$ is the upper critical field taken to be 80 T.\cite{puz95,tho96, leb96} Solving for the penetration depths we find $\lambda_{||} (T=0) = 1,760 \pm 60$ \AA \ and $\lambda_{\perp}(T=0) =  1.7 \pm 0.2\,\mu$m. Earlier results in near-zero applied field from imaged Josephson vortices\cite{kir98} found $\lambda_{\perp}(T=0) \approx  8 \,\mu$m , significantly larger than ours. However,  $\lambda_{||}$ and $\lambda_{\perp}$ from susceptibility on powders\cite{leb96, tho96, pan97} are consistent with our work. This comparison shows that there is no significant variation in the penetration depth for fields up to $H_0 = 6.4$ T indicating the minimal role of non-local effects due to Doppler contributions, also referred to as the Nonlinear Meissner Effect.\cite{xu95}    

	Absence of magnetic field shifts from O(2) NMR rules out the existence of static circulating orbital currents at the apical oxygen site in \HBCO.  However, discrepancy between detection of magnetic ordering by neutrons which are absent in NMR measurements is not unique to Hg1201.  For example antiferromagnetic ordering has been observed by neutron scattering in UPt$_3$\cite{aep88} but not detected in NMR measurements which was suggested to be a consequence of the  effects of thermal fluctuations and different time scales for these experiments.\cite{lee93, fom96} Roughly speaking these are 10$^{-11}$s for neutrons,\cite{fau06} 10$^{-8}$s for $\mu$SR,\cite{mac08} and 10$^{-6}$ for NMR frequency shifts. In the context of Hg1201 this was noted by Li et al. \cite{li11}

In summary, we have measured the temperature, field, and angular dependence of the narrow planar and apical \Ox\ NMR spectra in an underdoped Hg1201 single crystal.  There are no indications of static magnetic fields from circulating orbital currents proposed theoretically, within our accuracy of 0.4 G.  The  apical oxygen  spectra have a small temperature dependence above \Tc\ which has  axial symmetry, consistent with a direct dipolar field from the Cu$^{+2}$ site.

\textbf{ACKNOWLEDGMENTS}
Research was supported by the U.S. Department of Energy, Office of Basic Energy Sciences, Division of Materials Sciences and Engineering under Awards DE-FG02-05ER46248 (Northwestern University), DE-SC0006858 (University of Minnesota), and  the National High Magnetic Field Laboratory through the National Science Foundation and the State of Florida.

\bibliography{Bib.bib}
\end{document}